\documentclass[namedreferences]{solarphysics}
\usepackage[optionalrh]{spr-sola-addons} 
\usepackage{graphicx}                    
\usepackage{color}                       
\usepackage{url}                         
\usepackage{tabularx}

\begin{document}

\begin{article}

\begin{opening}

\title{Coronal Mass Ejections Observed at the Total Solar Eclipse on 13 November 2012}

%
\author{Yoichiro~\surname{Hanaoka}$^{1}$\sep
        Jun~\surname{Nakazawa}$^{2}$\sep
        Osamu~\surname{Ohgoe}$^{3}$\sep
        Yoshiaki~\surname{Sakai}$^{4}$\sep   
        Kazuo~\surname{Shiota}$^{3}$      
       }

%

%
  \institute{$^{1}$ National Astronomical Observatory of Japan, Mitaka, Tokyo, Japan
                     email: \url{hanaoka@solar.mtk.nao.ac.jp} \\ 
             $^{2}$ Yamanashi Cement Company, Kofu, Yamanashi, Japan \\
             $^{3}$ Solar Eclipse Information Center, Japan \\
             $^{4}$ Chiba Prefectural Kazusa High School, Kimitsu, Chiba, Japan
             }

\begin{abstract}
We carried out white-light observations of the total solar eclipse on 13 November 2012 at two sites, where the totality occurred 35 minutes apart.  We caught an ongoing coronal mass ejection (CME) and a pre-CME loop structure just before the eruption in the height range between 1--2 $R_\odot$.  The source region of CMEs was revealed to be in this height range, where the material and the magnetic field of CMEs were located before the eruption.  This height range includes the gap between the extreme ultraviolet observations of the low corona and the spaceborne white-light observations of the high corona, but the eclipse observation shows that this height range is essentially important to study the CME initiation.  The eclipse observation is basically just a snapshot of CMEs, but it indicates that future continuous observations of CMEs within this height range are promising.
\end{abstract}

%

\end{opening}

\section{Introduction}

Total solar eclipses provide us with rare chances to obtain data of the white-light corona with a wide dynamic range and a high signal-to-noise ratio from the solar limb to a couple of solar radii, and such observations are difficult to carry out with other methods.  In particular, recent observations and sophisticated data processing show coronal fine structures with superb visibility after the pioneering work by \inlinecite{Druckmuller2006}.  

On 13 November 2012, near the solar maximum of cycle 24, a total eclipse occurred.    At this eclipse, the Moon's umbral shadow passed from the Australian continent to the southern Pacific Ocean.  To take white-light images of the corona, we collaborated with participants of commercial eclipse tours, who had small telescopes/telephoto lenses and digital single-lens reflex (DSLR) cameras.  Although the weather conditions were not very good, observers at two sites, where the totality occurred 35 minutes apart, succeeded in obtaining high-quality white-light data of the eclipse with plenty of calibration data.  Data taken with various exposure times enabled us to produce a well-calibrated coronal image with a wide dynamic range and a high signal-to-noise ratio.

Owing to the high frequency of the occurrence of coronal mass ejections (CMEs) under high solar activity, manifestations of CMEs were observed at the eclipse in the height range of 1--2 $R_\odot$ in the corona.  One of them is a CME started between the eclipse observations at the two sites.  It was noticed as an ongoing CME by a number of observers \cite{Ohgoe2013}.  Another one is a coronal loop structure, which developed into a CME soon after the eclipse.

These CMEs are also observed in extreme ultraviolet (EUV) at low altitude ($<$ 1.5 $R_\odot$) by the Atmospheric Imaging Assembly (AIA) on board the Solar Dynamics Observatory (SDO) and in visible continuum at high altitude (mainly $>$ 2 $R_\odot$) by the Large Angle and Spectrometric Coronagraph (LASCO) on board the Solar and Heliospheric Observatory (SOHO) as many other cases of recent CMEs.  Note that there is a blank gap in the height coverage of these spaceborne instruments.  
Generally, the EUV observations show the bottom part of the structural evolutions of the corona due to CMEs.  On the other hand, the spaceborne coronagraph observations show already ejected material in the high corona.
Therefore, the CME material before the eruption is probably located in the height gap, and the build-up of the magnetic field structure, which eventually causes a CME, probably also takes place there.  

The Sun Earth Connection Coronal and Heliospheric Investigation (SECCHI) of the Solar Terrestrial Relations Observatory (STEREO) also observed the above CMEs from different directions.  SECCHI has overlapping height coverage with the Extreme UltraViolet Imager (EUVI, $<1.7 R_\odot$) and a coronagraph (COR1 $>1.4 R_\odot$), but the discontinuity in the observing wavelength makes it difficult to catch coronal structures ranging over 1--2 $R_\odot$.  An EUV imager, the Sun Watcher using Active Pixel System detector and image Processing (SWAP) of the Project for On-Board Autonomy-2 (PROBA2) has a field of view covering up to nearly 2 $R_\odot$, but the upper layer of the corona in the SWAP field of view is too faint to obtain data with a sufficient signal-to-noise ratio.

The ground-based coronagraphs have observed CMEs in the height range including the height gap.  In particular, using the coronagraph of the Mauna Loa Solar Observatory, \inlinecite{StCyr1999} measured the acceleration of CMEs, which mainly occur in the corona lower than the spaceborne coronagraph height, and \inlinecite{Gibson2006} showed the linkage between coronal cavities in the low corona and CMEs.  
However, the ground-based coronagraph observations are disturbed by the bright sky, and its signal-to-noise ratio is limited.
In space, the C1 coronagraph of LASCO, which had a capacity to observe low height down to 1.1 $R_\odot$, observed CMEs in the height gap ({\it e.g.} \opencite{Srivastava1999}; \opencite{Plunkett2000}), but LASCO C1 ended its observation as early as in 1998.  

To understand CMEs further, observations of the height range 1--2 $R_\odot$ with a high signal-to-noise ratio are indispensable.  The observation of the 13 November 2012 eclipse is a good example of such observations.  Eclipse observations are basically a `snapshot' of CMEs, but they can show what is expected in the observations of the corona in the height gap, and to what aspect of the CME science the observations in the height gap will contribute.  Thus they will be useful to scheme out future coronal observations, which have appropriate coverage of the height range to study CMEs.  

The observation and the reduction of the eclipse data are described in Section 2.  The analysis of the eclipse data and the comparison with the data taken with other instruments are given in Section 3.  In Section 4, we present summary and discussion.

\section{Eclipse Observation and Data Reduction}

The observational data of the 13 November 2012 eclipse were obtained at two sites.   The first one is Mareeba, Queensland, Australia (16S145E, the maximum eclipse 20:39:36UT, the duration of the totality 1 m 40 s, the altitude of the Sun 14$^\circ$).  The total eclipse umbra left the Australian continent in early morning, and then it traveled on the Pacific Ocean to the last.  After 35 minutes, the eclipse was observed at the second site, on the commercial tour ship Pacific Venus (30S173E, the maximum eclipse 21:14:49UT, the duration of the totality 3 m 20 s, the altitude of the Sun 48$^\circ$), cruising north of New Zealand on the southern Pacific Ocean.
We successfully obtained several sets of data, which are appropriate for the photometric use.  Each of these data sets contains 25 or more images of the corona taken during a totality with various exposure times using telescopes/telephoto lenses with apertures of 5--10 cm. 
For the analysis of CMEs, we adopted one of the data sets for each site; the one at Mareeba was taken with a Takahashi FS-102 refractor (102 mm aperture, 820 mm focal length) + a Canon EOS 7D camera (operated by co-author YS); the other one on Pacific Venus was taken with a BORG 100ED refractor (100 mm aperture, 512 mm synthesized focal length) + a Canon EOS KissX3 camera (operated by co-author OO).

These data were reduced using basically the same method as that used in the 2008 and 2009 eclipses \cite{Hanaoka2012}.  Dark current and flat field corrections were applied for each image, and non-linear response of the analog-to-digital conversion for each camera was corrected.  The images were normalized by the exposure time, and the recorded signals were converted to the amount of the incident light per second.  These normalized images were stacked to produce a single image with a high signal-to-noise ratio and a wide dynamic range.  Since the position of the sun is not fixed on the images (particularly those taken on the rolling vessel), the images were aligned based on the coronal streamer structures.  Properly exposed parts of the images were used to stack.
Then the apparent brightness on the images was converted to the brightness relative to the solar disk by measuring the brightness of the partially eclipsed sun and the sun before/after the first/fourth contacts taken with neutral density filters.  
The image scale and the direction of the north were determined on one of the stacked images using stars whose positions are taken from the Tycho-2 catalogue \cite{Hog2000}.  The scale and the orientation of the other images were determined by comparison with the reference image.  The correction for the atmospheric refraction was applied to the images taken at the first totality (Mareeba), because the altitude of the Sun was only $14^{\circ}$ then.
The stacked images of the white-light corona for the two totalities are shown in Figures 1a and 1b with enhanced contrast for coronal fine structures.  

The results of the analysis of these images are presented in Section 3, but here we mention the total brightness of the white-light corona at the 2013 eclipse.
The total brightness (1.03 -- 6 $R_\odot$) is derived to be 0.85 in unit of $10^{-6} L_\odot$, where $L_\odot$ is the total brightness of the solar disk.  This value is about two times those at the 2008 and 2009 eclipses \cite{Hanaoka2012} occurring during the deep solar minimum between cycles 23 and 24, but somewhat smaller than the average coronal brightness measured near the solar maxima in the twentieth century \cite{Rusin2000}.  This is consistent with the fact that the sunspot relative number in cycle 24 is lower than the typical solar maximum value.

%
\begin{figure} 
\centerline{\includegraphics[width=1\textwidth,clip=]{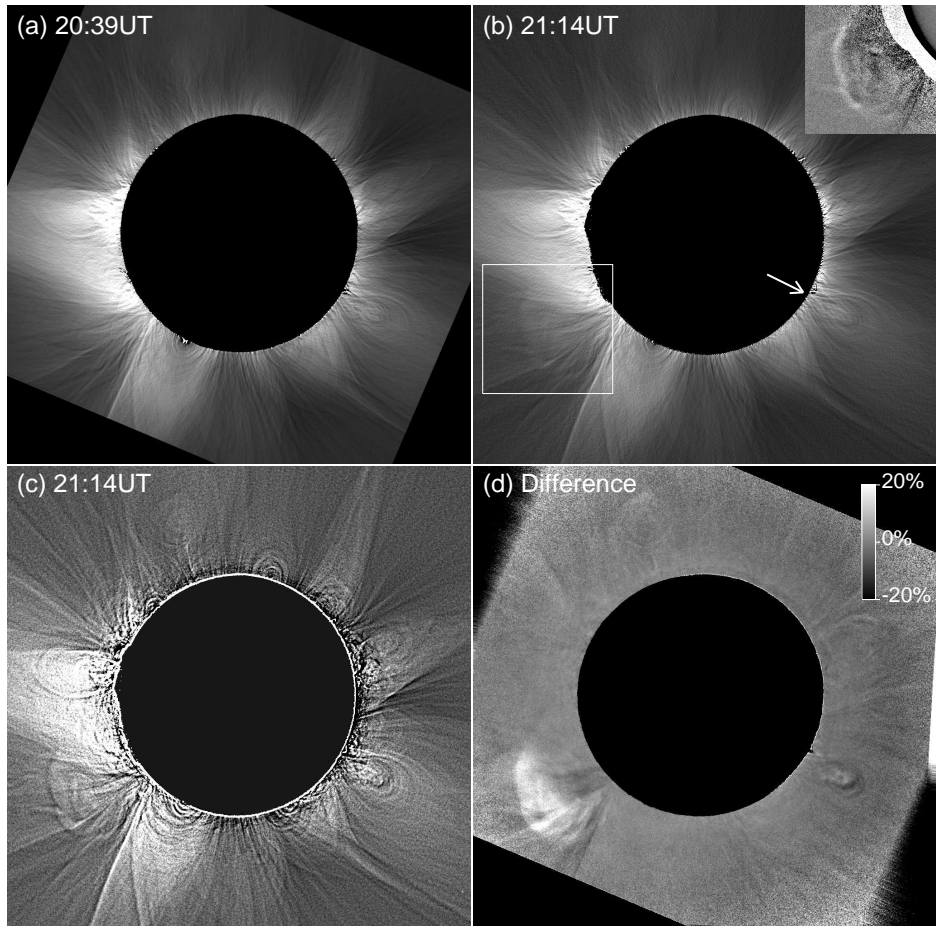}}
\caption{(a)(b) Stacked images of the white-light corona at (a) 20:39 UT (b) 21:14 UT of the 13 November 2012 eclipse.  Fine structures of the corona are enhanced.  The inset at the upper right corner in panel b is a difference image between two images taken at the totality around 21:14 UT showing an ongoing CME.  A box in panel b shows its position.  An arrow points to a loop structure developing into another CME.  (c) The image at 21:14 UT with enhanced loop structures.  (d) Relative difference image between the images at 20:39 UT and 21:14 UT.
The solar north is to the top.
}
\end{figure}

\section{Data Analysis of the CMEs}

\subsection{Overview of the Results from the Eclipse Observation}

The stacked image of the corona taken at the first totality at 20:39UT and that at the second totality at 21:14UT are shown in Figure 1 (panels a and b), along with the image at 21:14UT showing enhanced loop structures (panel c) and a relative difference image between 20:39 and 21:14 (panel d).  There is a noticeable difference between the images taken at the two totalities; a bubble-like structure at the east limb is seen only in the 21:14 image (in the box in panel b), and it is also clearly seen in Figure 1c.  This is an ongoing CME started between the first and the second totalities.  The upper-right inset in Figure 1b shows a difference image of the CME area between two individual images taken two minutes apart (21:14:13 and 21:16:12) during the second totality.  Bright edge and slight dimming inside the CME bubble seen in this difference image show the expansion of the CME bubble over two minutes, as \inlinecite{Ohgoe2013} points out.  Since 1860 \cite{Eddy1974}, before the nature of CMEs was understood, CMEs have been captured at total solar eclipses, and the detection of CMEs at eclipses is known to be not rare \cite{Webb1995}.  However, this seems to be the first observation of the expansion of a CME bubble during a totality.  
Relative brightness change between the two totalities at 20:39 and 21:14, $(I_{21:14}-I_{20:39})/I_{20:39}$, is shown in Figure 1d.  Obvious bubble-like brightness increase due to the CME can be seen.  Below the CME bubble, we can find coronal dimming as well.

On the other hand, at the west limb we can find a loop structure like a hot-air balloon (inverted teardrop shape) both in Figures 1a and 1b (designated by an arrow in panel b), and the difference image in Figure 1d shows that the loop structure changed slightly between the totalities.  This is the loop structure developed to a CME observed by LASCO after the eclipse.

\subsection{The CME at the East Limb}

%
\begin{figure} 
\centerline{\includegraphics[width=1\textwidth,clip=]{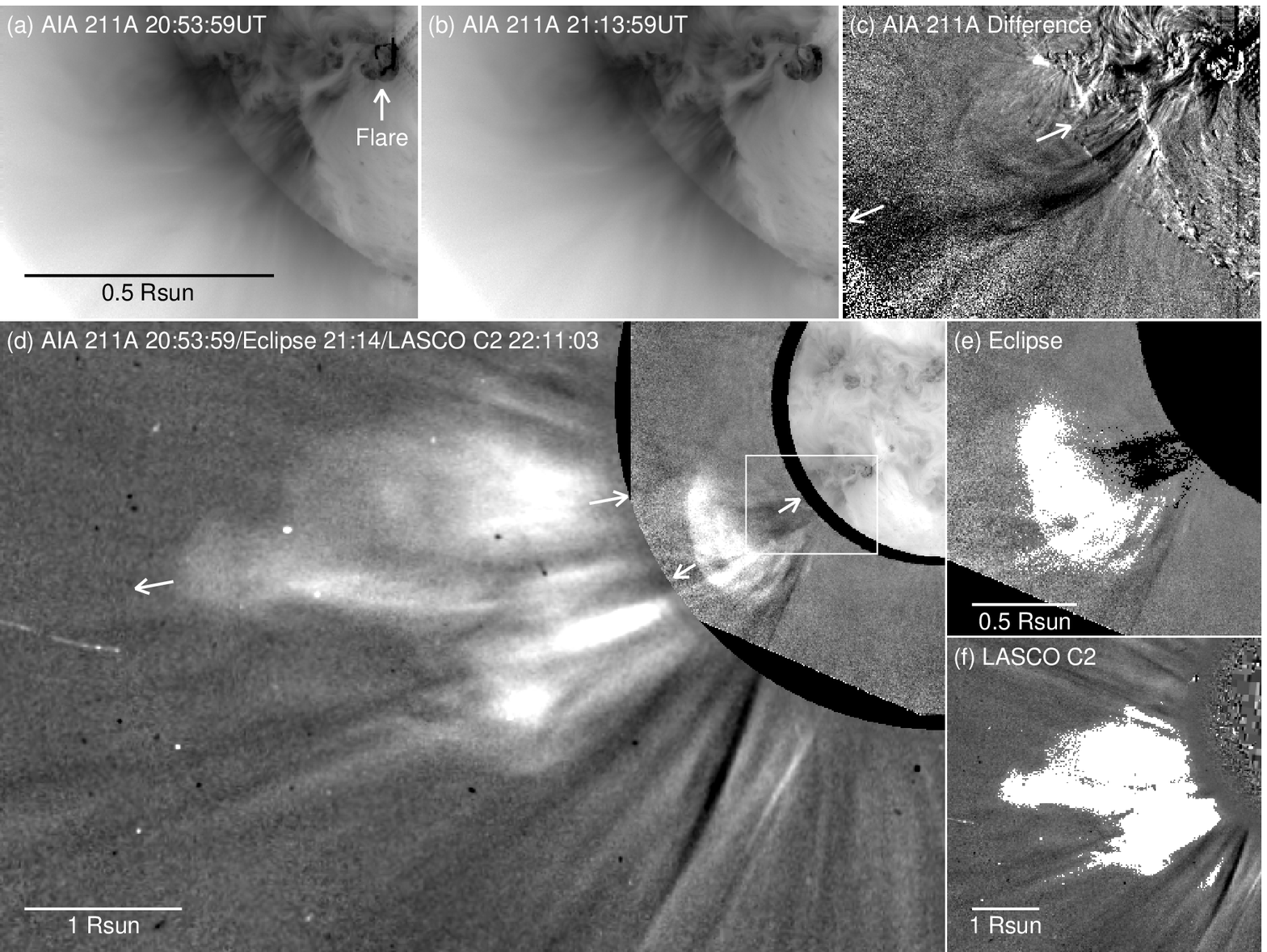}}
\caption{
(a)--(c) AIA Fe {\sc xiv} 211 \AA\ images related to the CME at the east limb. (a) At the flare maximum, (b) at the second totality of the eclipse, and (c) their relative difference image (panels a and b are shown in negatives).  (d) Aligned picture of the CME at the east limb consisting of the AIA 211 \AA\ image (negative) at the flare (20:53:59) ($< 1 R_\odot$), the eclipse corona difference image (1 -- 2.1 $R_\odot$), and the LASCO relative difference image between 22:11:03 and 20:39:45 ($> 2.1 R_\odot$).  A box in panel d shows the field of view of panels a--c.  Arrows in panels c and d show the location of the strips used for the time-slice chart in Figure 3.  (e)(f) Pixels used to calculate the total brightness of the CME (white) and the dimming (black) in (e) the eclipse image and (f) the LASCO C2 image.
}
\end{figure}

%
\begin{figure} 
\centerline{\includegraphics[width=0.5\textwidth,clip=]{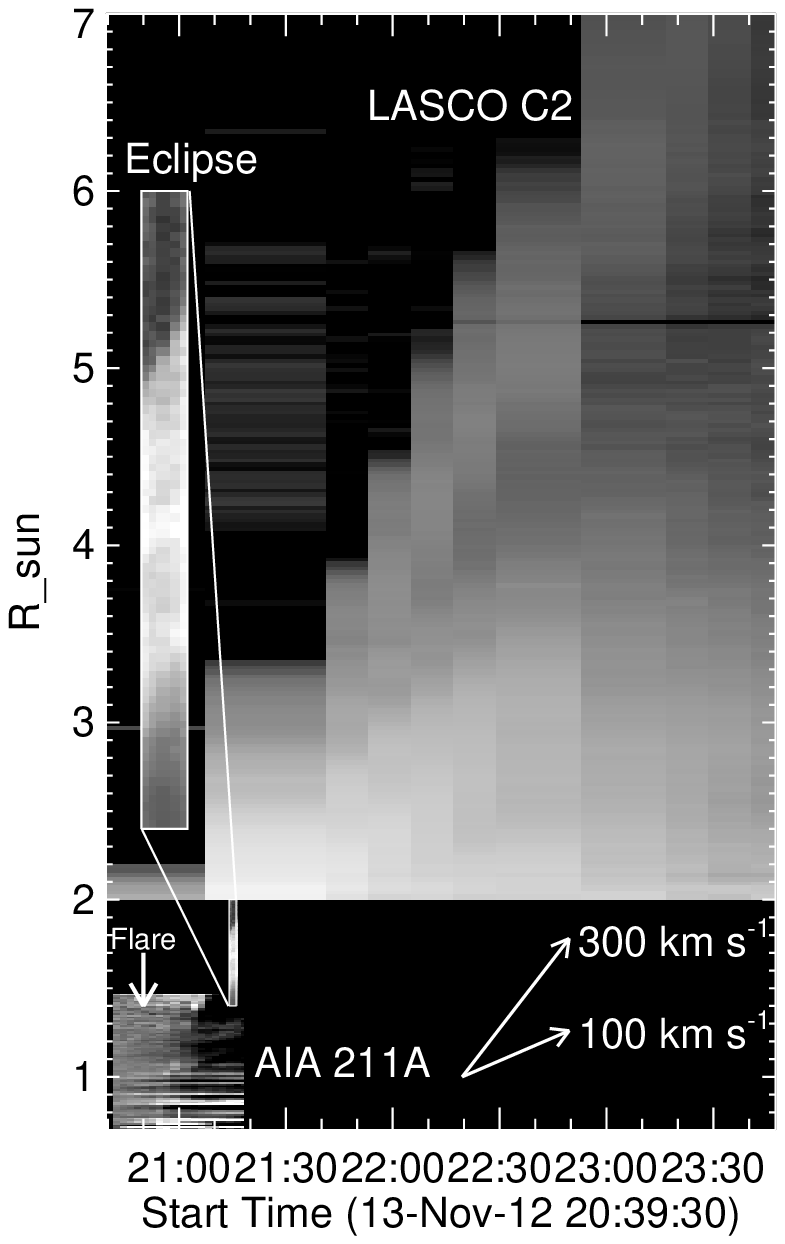}}
\caption{
Time slice chart of the strips designated in Figures 2c and 2d, showing the temporal evolution of the brightness along the slice in base-difference images of AIA 211 \AA , eclipse (at the second totality), and LASCO C2.  Because the eclipse images covers only two minutes, a blowup chart is also shown.  The onset of the flare, 20:50, is indicated.
}
\end{figure}

Figure 2 shows EUV (Fe {\sc xiv} 211 \AA ) images taken with SDO/AIA and white-light data taken with the SOHO/LASCO C2 coronagraph along with the eclipse data.  Between the eclipse observations at the two sites, an M2.8 flare occurred in region NOAA 11613 at S22E33 (according to the flare list of the Geostationary Operational Environmental Satellite [GOES], start: 20:50, max: 20:54, and end: 20:57), and its location is designated in Figure 2a.  The 211 \AA\ image in Figure 2a at the flare maximum (20:53) and that in Figure 2b near the second totality of the eclipse (21:13) show no obvious difference, but their relative change, $(I_{21:13}-I_{20:53})/I_{20:53}$, in Figure 2c shows a dimming region rooted in the flare location.  Figure 2d shows an aligned picture of the AIA 211 \AA\ image at the flare ($< 1 R_\odot$), the eclipse corona difference image showing the CME (1--2.1 $R_\odot$), and the LASCO difference image ($> 2.1 R_\odot$) between 22:11 and 20:39, $(I_{22:11}-I_{20:39})/ I_{20:39}$, showing the CME fully entering the LASCO C2 field of view one hour after the eclipse observation.  The dimming seen in the eclipse image below the CME bubble can be considered as an extension of that seen in the AIA difference image, of which the field of view is shown by a box in Figure 2d.  Therefore, the CME bubble and the flare are connected by the dimming, and the flare is considered to have caused the CME and the dimming.

In the eclipse difference image in Figure 2e, pixels whose brightness increase is larger than a certain threshold are picked up so that they cover the CME bubble, and they are filled with white.  On the other hand, the pixels covering the dimming area are filled with black.  The total brightness increase of the white pixels is calculated to be $5.7\times 10^{-10} L_{\odot}$, and the total brightness decrease of the black pixels is $4.9\times 10^{-10} L_{\odot}$.  Considering that part of the dimming in the eclipse image is covered by the moon (see Figure 2c), we can conclude that the amount of brightening is consistent with that of the dimming, and therefore, the dimming area is deemed to be the source of the CME material.  This means that the CME source region in the height range of 1--2 $R_\odot$ was captured by the eclipse observation, which fills the gap between the AIA and LASCO observations.
The flare looks like a compact one, but it triggered an eruption of coronal material ranging beyond the AIA field of view.  
The pre-CME images of the eclipse (Figure 1a) and AIA (Figure 2a) show that the streamer structure in the source region of this CME seemed to be mostly open, and it did not show a characteristic structure seen in usual CME source regions such as a helmet streamer or a coronal cavity.  However, the observation in the height range of 1--2 $R_\odot$ shows that a CME can occur in such a region.

The CME seen in the eclipse image is apparently about 1.5--1.8 $R_\odot$ away from the disk center.  Because the location of the flare is about $40^\circ$ from the disk center, the actual radial distance of the CME from the disk center is estimated to be around 2.7 $R_\odot$.  Considering this distance, we estimated the mass of the CME material to be $7\times 10^{14}$ g based on the Thomson-scattering efficiency.  The typical mass of a CME estimated so far (see {\it e.g.} \opencite{Webb2012}) is on the order of $10^{15}$ g.  Therefore, this is a small CME.

In Figure 2f showing the CME area in the LASCO difference image, the pixels covering the CME area are filled with white.  Total brightness of the LASCO CME is $3.0\times 10^{-10} L_{\odot}$, namely, roughly half of that at the eclipse.  At 22:11, the CME was located about 1.5 times farther than that at the eclipse time, 21:14, from the disk center.  
The 1.5 times difference makes the solid angle of the solar disk seen from the CME (the source of the light scattered by the CME) about half.  Therefore, the CME brightness at the eclipse and that in the LASCO image indicate the amounts of CME material seen in these observations to be consistent with each other. 

Figure 3 shows temporal evolution of the brightness along the slices across the CME in the AIA, eclipse, and LASCO images.  Paths connecting the arrows in Figures 2c and 2d indicate the positions of the slices.  The AIA 211 \AA\ images were taken approximately every two minutes, and the LASCO C2 images were taken approximately every 12 minutes.  There are five individual images of the eclipse taken with a proper exposure time for the CME with an interval of about 30 seconds.  Base differences of these images are shown in Figure 3.
In the AIA time slice, we can find expansion of the dimming, which started roughly at the flare onset, and the velocity of its leading edge is about 600 km s$^{-1}$.  The eclipse data show the upward motion of the CME bubble with the velocity of the leading edge of about 360 km s$^{-1}$.  The LASCO time slice shows the velocity of the leading edge of about 310 km s$^{-1}$.  These velocities are somewhat smaller than the average velocity of CMEs \cite{Webb2012}.  However, because the location of the flare was 40$^\circ$ from the disk center, the actual velocity is presumed to be about 500 km s$^{-1}$, which is not much different from the average.

Tracing back the motion of the CME bubble extending 1.5--1.8 $R_\odot$ at the eclipse time to the flare onset, we can estimate the apparent height of the main body of the CME to be 0.9--1.2 $R_\odot$.  With the correction of the projection effect, the actual height of the mail body of the CME at its onset is estimated to be 1.4--1.9 $R_\odot$.  This means that the major part of the CME source region extends to high above the compact flare region.

STEREO also has visible light coronagraphs; Figure 6d shows the CME taken by one of them, SECCHI/COR1 of STEREO-B, which was located 124 deg behind the earth at the eclipse.  The CME at the east limb seen from the earth was seen at the west limb from STEREO-B, and its velocity is estimated to be about 340 km s$^{-1}$.  The velocities based on the LASCO and STEREO observations indicate that the direction of the CME movement is about 30$^\circ$ toward the earth from the sky plane seen from the earth.

\subsection{The CME at the West Limb}

%
\begin{figure} 
\centerline{\includegraphics[width=1\textwidth,clip=]{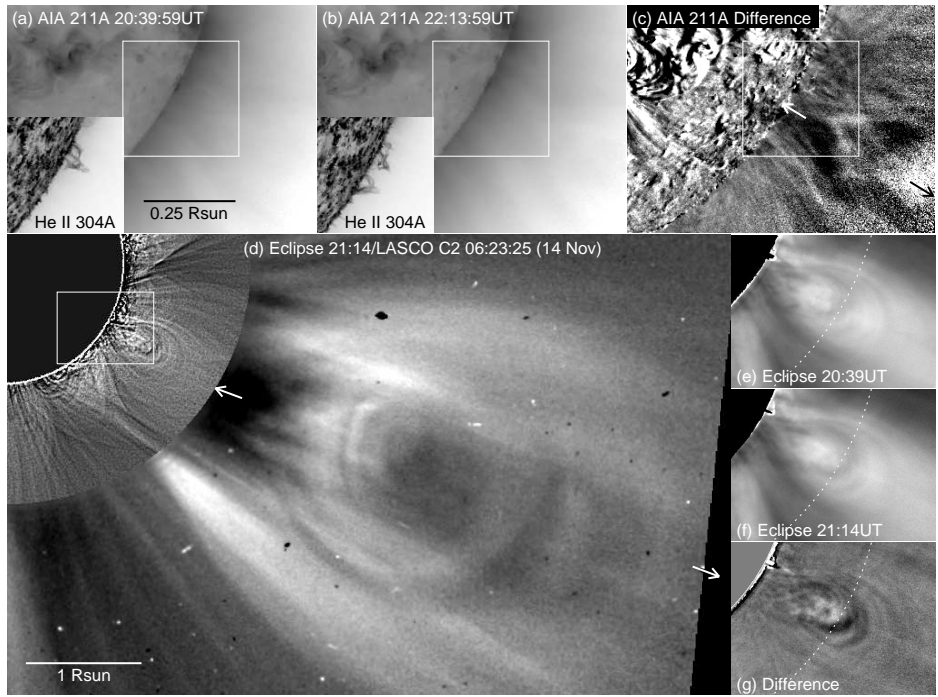}}
\caption{
(a)--(c) AIA Fe {\sc xiv} 211 \AA\ images related to the CME at the west limb. (a) At the pre-CME phase, (b) one hour after the second eclipse observation, and (c) their relative difference image (panels a and b are shown in negatives).  The insets in panels a and b are AIA He {\sc ii} 304 \AA\ images taken eight seconds after the 211 \AA\ pictures.  The position of the 304 \AA\ images is shown by boxes in panels a--c.
(d) Aligned picture of the CME at the west limb consisting of the eclipse corona difference image showing the hot-air balloon loop structure ($< 2.1 R_\odot$), and the LASCO relative difference image between 20:39:45 (13 November) and 06:23:25 (14 November) ($> 2.1 R_\odot$).  A box in panel d shows the field of view of panels a--c.  Arrows in panels c and d show the location of the strips used for the time-slice chart in Figure 5.  (e)--(g) Blow-ups of  the hot-air balloon loop structure in the eclipse images taken at the two totalities and their difference.  Dotted-line circles show the position of 1.45 $R_\odot$.
}
\end{figure}

%
\begin{figure} 
\centerline{\includegraphics[width=1\textwidth,clip=]{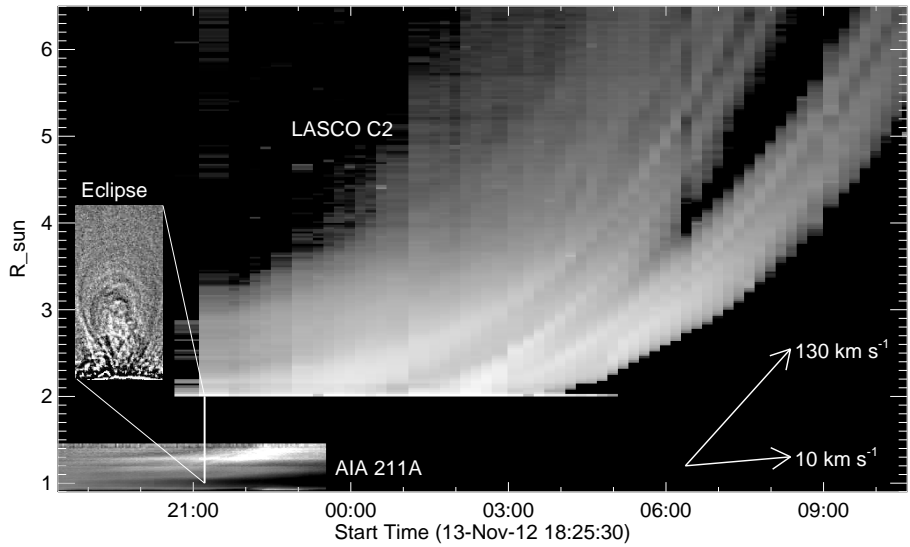}}
\caption{
Time-slice chart of the strips designated in Figures 4c and 4d, showing the temporal evolution of the brightness along the slice in base-difference images of AIA 211 \AA\ and LASCO C2.  The time of the second eclipse totality, 21:14, is shown by a vertical bar, and an image of the loop structure seen at the eclipse, covering the height range of 1--2 $R_\odot$, is also shown.
}
\end{figure}

%
\begin{figure} 
\centerline{\includegraphics[width=0.8\textwidth,clip=]{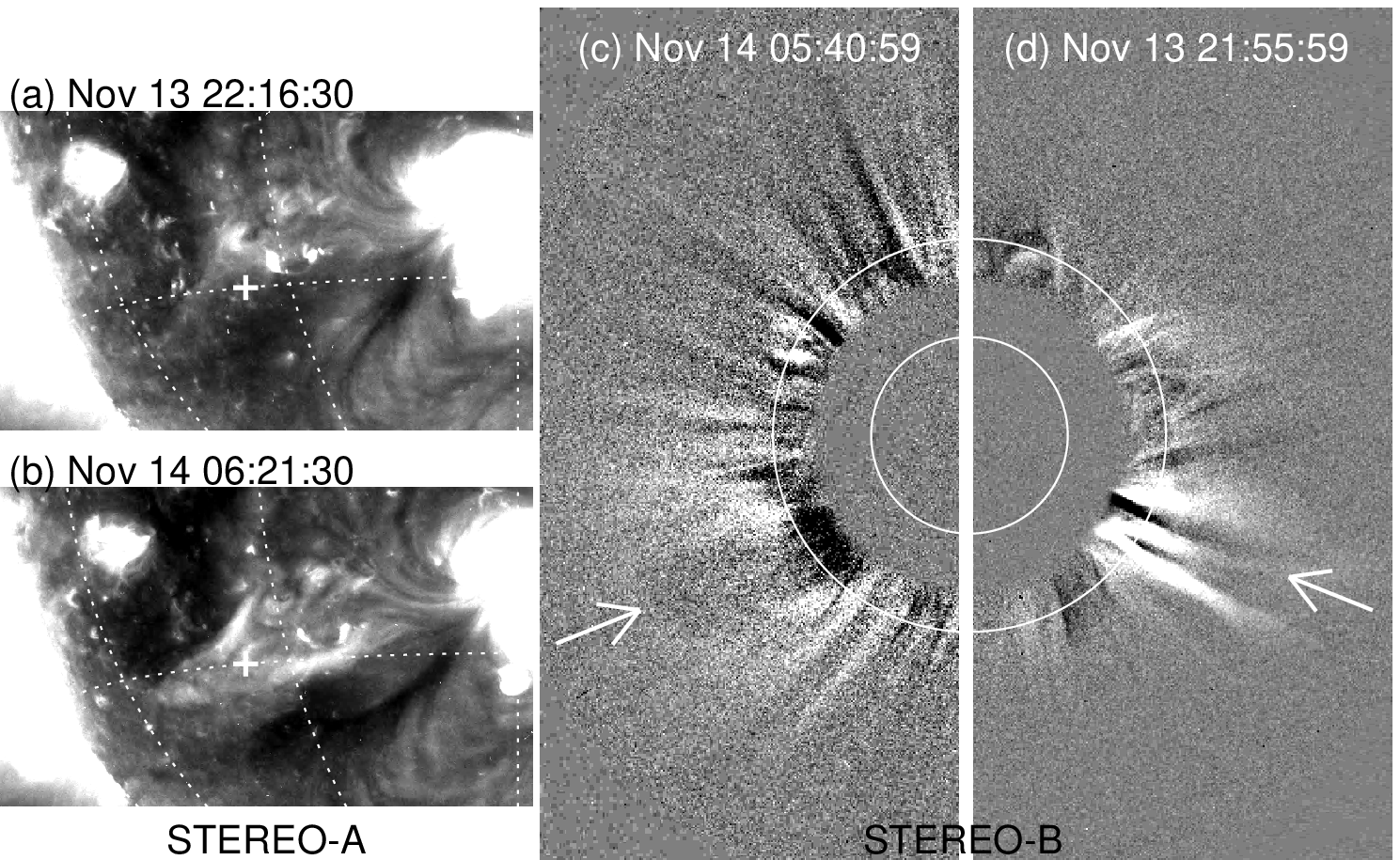}}
\caption{
(a)(b) Fe {\sc xiv} 284 \AA\ images of the corona below the CME shown in Figure 4 taken by STEREO-A/SECCHI/EUVI. The footpoint of the hot-air balloon loop structure is marked with a plus sign in each panel. (c)(d) White-light difference images taken by STEREO-B/SECCHI/COR1 showing CMEs (designated by arrows).  Panel c shows the CME in Figure 4, and panel d shows the other CME in Figure 2.
}
\end{figure}

The CME at the west limb slowly started to erupt after the eclipse, and its rise was seen for several hours.  Figure 4 shows AIA 211 \AA\ and LASCO C2 images of this CME along with the eclipse data.  
In Figures 4a--c, a pre-CME 211 \AA\ image at the first eclipse time (20:39), that at one hour after the second eclipse observation (22:13), and their relative change, $(I_{22:13}-I_{20:39})/I_{20:39}$, are shown.  
Figure 4d is an aligned picture of the eclipse corona ($< 2.1 R_\odot$) with enhanced contrast in the pre-CME phase at 21:14 UT and a LASCO difference image ($> 2.1 R_\odot$) between 20:39 (13 November) and 06:23 (14 November) showing the ongoing CME.

Figures 4a and 4b do not show remarkable change, but in the difference image in Figure 4c we can find a structure corresponding to the lower half of the hot-air balloon seen in the eclipse image in Figure 4d (the field of view of Figure 4c is shown by a box in panel d).  
The LASCO image in Figure 4d shows circular (possibly helical) loop structures.  Based on the morphological similarity, it is concluded that the hot-air balloon loops in the eclipse image erupted upward and became the CME.  Below the CME, we can find a post-eruptive arcade.  Figures 6a and 6b show Fe {\sc xv} 284 \AA\ images in the pre-CME phase (22:16) and during the CME (06:21) taken by EUVI of SECCHI on board STEREO-A, which was 127$^\circ$ ahead of the Earth at the eclipse.  A plus sign in each panel shows the footpoint of the hot-air balloon loops, which is assumed just on the limb seen from the Earth.  Figure 6b taken around the same time as the LASCO image in Figure 4d shows that a faint post-eruptive arcade was developed below the ongoing CME.  As a whole, this event is a standard CSHKP (acronym of Carmichel, Sturrock, Hirayama, Kopp, and Pneuman; see \opencite{Svestka1992}) type one, of which the whole image combining a CME and arcade loops were presented by, e.g., \inlinecite{Lin2004}.
Figures 4e--g show blowups of the pre-CME loop structure in the eclipse images taken at the two totalities and their difference image.  Unlike the eclipse picture in panel d, panels e and f are shown without contrast enhancement except that the brightness is normalized by the average brightness at each height.  As seen in these pictures, the pre-CME loop structure was slowly ascending (compare the positions of the central core and the cavity above it with dotted-line circles at 1.45 $R_\odot$) with the speed of about 10 km s$^{-1}$. 

The above results indicate that only the eclipse observation shows the whole appearance of the CME source region, which provided erupting magnetic field and material of the CME.  It extends to 2 $R_\odot$ or higher, and its slow rise is considered to imply that the magnetic field in the source region was evolving from a stable to an unstable state leading to the eruption.
Therefore, the eclipse observation, which again fills the gap between the AIA and the LASCO observations, indicates that the height range of 1--2 $R_\odot$ is crucially important to study the initiation of the CME.

The time evolution of the CME seen in the AIA and the LASCO C2 images is shown in Figure 5, which shows the time slices along the strips shown in Figures 4c and 4d by arrows.  In the AIA time slice, we can find slowly ascending structures, of which the velocity is about 8 km s$^{-1}$.  As mentioned above, the eclipse images show the ascending motion of about 10 km s$^{-1}$.  The LASCO data clearly show the acceleration of the CME to the velocity of 130 km s$^{-1}$ at the height of 4--5 $R_\odot$ after the eclipse.  
Although this erupting speed is quite low compared to that of the typical CMEs \cite{Webb2012}, such a low speed is not rare, and is particularly common in stealth CMEs \cite{Ma2010}, which show only weak surface association with CMEs.
In contrast to the CME at the east limb described in Section 3.2, the hot-air balloon loop structure showed a gradual evolution, which finally resulted in the eruption.  This fact suggests that the continuous magnetic evolution in the photosphere brought about the instability of the magnetic field (see \opencite{Forbes2006} and \opencite{Magara2012} for the discussion on the evolution leading to the onset of CMEs).
The fact that there is a hot-air balloon loop structure in the slowly developing phase of the CME suggests that the coronal loops, which are about to erupt, have a shape of a flux rope rather than a common helmet streamer.  Such a magnetic field structure, on the verge of erupting, has been studied by model calculations (e.g. \opencite{Linker2003}; \opencite{Lin2004}; \opencite{Wu1997}), and the observational results, which show the structure of the source region of CMEs, will contribute to advancing the building of CME models.

The combination of slow rise of a flux rope-like structure and a later CME was also studied by \inlinecite{Regnier2011}.  They only observed U-shape loops under a coronal cavity in images taken by AIA, but the observation with a wider field of view such as the eclipse observation would show the complete loops around the cavity.  In the case studied by \inlinecite{Regnier2011}, there was a prominence at the bottom of the U-shaped loops.
Generally, a flux rope suggests the existence of a prominence at its bottom (Forbes 2006).
The inset in Figure 4a, a He {\sc ii} 304 \AA\ image taken by AIA at 21:14, shows a prominence.  It might be guessed to be related to the hot-air balloon loop structure, but actually, it is located north of the footpoint of the loop structure (see the box showing the position of the He {\sc ii} image in Figure 4c; the white arrow is located at the footpoint of the loop structure).  Therefore, this prominence is not related to the loop structure.  The inset in Figure 4b, another He {\sc ii} 304 \AA\ image at 22:14, shows that another prominence appeared at the footpoint of the loop structure.  This prominence disappeared soon, and it did not seem to erupt with the CME.  It is not clear if there is a prominence related to the hot-air balloon loop structure.

Figure 6c shows the CME taken by SECCHI/COR1 of STEREO-B.  The CME was seen at the east limb from STEREO-B.  The velocity of the CME observed by STEREO-B is about 100 km s$^{-1}$.  On the basis of the LASCO and STEREO observations, the direction of the CME movement is presumed to be 15--20$^\circ$ away from the sky plane.

\section{Summary and Discussion}

The observation of the eclipse on 13 November 2013 showed two CME-related coronal structures.  

One of them was observed as an ongoing CME.  The CME was triggered by a flare and the ejection started abruptly.  Though the flare was a rather compact one, the eclipse observation shows that the source region of this CME extended outside of the AIA field of view.  There was no helmet streamer or coronal cavity in the source region.  The eclipse observation in the height range of 1--2 $R_\odot$ shows that a CME can occur in such a region.

The other one is a pre-CME flux rope-like loop structure extending from the solar limb to over 2 $R_\odot$, and it developed into a standard CSHKP-type event consisting of a CME and a post-eruptive arcade after the eclipse.  Contrary to the CME described above, it was evolving very slowly during the eclipse observation, and eventually it erupted, probably without any abrupt trigger.  The loop structure observed at the eclipse corresponds to the CME source region and it shows the coronal structure just before the eruption.

These results mean that observations with a high signal-to-noise ratio covering the blank gap of the recent space observations provide information about the source region of CME material and magnetic field.
Therefore, such observations are essential to an understanding of the CME initiation.  Eclipse observations cover the height range 1--2 $R_\odot$ and achieve the high signal-to-noise ratio by widening the dynamic range, but these are basically snapshot-type observations.  Therefore, continuous observations are desirable, and they will provide information enabling us to predict CMEs.

Then what kind of instrument does accomplish such observations?  Regarding the height coverage, a Lyot-type spaceborne coronagraph, which is similar to the SECCHI white-light instrument with an occulting disk being as small as possible, is suitable to observe the height range 1--2 $R_\odot$ in white light.  Such a coronagraph will also be suitable for observations using visible light emission lines of the corona as LASCO C1.  For the visible light observations, there is another idea for observing eclipses from a moon-orbiting spacecraft using the moon as an occulter \cite{Habbal2013}.
An EUV imager with a wide field of view is also a candidate.
However, the dynamic range should meet the requirement as well.  In Figures 4e and 4f showing the pre-CME loop structure, the brightness is normalized by the average brightness at each height.  The average brightness of the corona at 1.1 $R_\odot$, for instance, is a hundred times as bright as that at 2.1 $R_\odot$.  Tangential brightness contrast of the individual loops seen in Figures 4e and 4f is several percent.  These facts mean that to resolve loop structures in the CME source region with a sufficiently high signal-to-noise ratio, the dynamic range of images is required to be at least several thousand.  Since this is difficult to achieve by means of a single exposure, a combination of various exposure times is required, as in the case of the eclipse observations.  Furthermore, observations with coronal emission lines require wider dynamic ranges.  Our eclipse observation was done with white light, namely the Thomson scattering light, of which brightness is proportional to the electron density.  However, the brightness of the coronal emission lines has stronger dependence on the electron density.

The eclipse observation revealed that continuous observations of the corona including the height range 1--2 $R_\odot$ with a wide dynamic range are promising, and that they will contribute to studying the initial phase of the CME very much, even though there are various difficulties in achieving such observations.  We anticipate that such observations will be regularly carried out in future.

%

%
\begin{acks}
One of the authors (YH) had appealed to the participants of commercial eclipse tours to take scientific data at the eclipse in 2012.  He thanks all the people who responded to the appeal and expressed their intention to try to take the data, though not all of them succeeded in the observation.
The SOHO/LASCO data used here are produced by a consortium of the Naval Research Laboratory (USA), Max-Planck-Institut f\"ur Aeronomie (Germany), Laboratoire d'Astronomie Spatiale (France), and the University of Birmingham (UK). SOHO is a project of international cooperation between ESA and NASA.
The AIA data used here are provided courtesy of NASA/SDO and the AIA science team. 
The STEREO/SECCHI data used here are produced by an international consortium: NRL, LMSAL, NASA, GSFC (USA); RAL (UK); MPS (Germany); CSL (Belgium); and IOTA, IAS (France).
\end{acks}

%
%
%

\end{article} 
\end{document}